\documentstyle[12pt,epsf]{article}
\catcode`\@=11 \@addtoreset{equation}{section}  
\renewcommand{\theequation}                     
         {1 .\arabic{equation}}   
\title{The Affine Gauge Theory in the Quantum Phase Space CP(N-1)}
\author{P. Leifer}
\date{Cathedra of Informatics, Crimea State Engineering and Pedagogical University, \\ 
21 Sevastopolskaya st., 95015 Simferopol, Crimea, Ukraine} 
\begin{document}
\maketitle

The ordinary linear quantum theory predicts the quantum correlations at any distance (the universal superposition principle). It creates the decoherence problem since quantum interactions entangle states into non-separable combination. On the other hand the linear quantum theory prevents the existence of the localizable solutions, and after all, leads to the divergences problem in the quantum field theory. In order to overcome these difficulties the non-perturbative nonlinearity originated by the curvature of the compact quantum phase space (QPS) has been used. 
\vskip 0.1cm
Non-linearity in quantum theory has been invoked in order to build the objective quantum theory and to prevent the unlimited spread out of the observable fields by the gravitational self-potential \cite{Jones1,Jones2}. But Newtonian quantum gravity in the present form is not effective for the shaping wave-packets of elementary particle size since the characteristic scale of the ground-state wave-packet obtained from the gravitational Schr\"odinger equation for nucleon masses is around $10^{23} m$ \cite{Jones2}. 

There is a different group of works make accent on the formulation of {\it the standard quantum mechanics} in QPS represented by the complex projective Hilbert space $CP(N-1)$
\cite{CMP,Hughston1,Hughston2,Ashtekar,AdHr}.  
I think, however, that consistent and prolific theory based on such QPS should be connected with serious deviations from the standard quantum scheme. Such modification must, of course, preserve all achievements of de~Broglie-Heisenberg-Schr\"odinger-Dirac linear theory by a natural way. One may think about attempts to establish a deductive approach to the quantum theory. 

{\bf General part.}---
The classical field theory treats the `fields' as force functions of the space-times coordinates with pointwise action on the probe particles. It is assumed this picture has clear macroscopic sense. Nevertheless, the classical field theory is contradictable since it is impossible to describe the stable charged particles in its own framework. Quantum particle like electron is pointwise but it is `wrapped' in so-called de~Broglie-Schr\"odinger fields of probability. 
Quantum field theory (QFT) uses same classical space-time coordinates as `index' whereas the fields are operators acting in some Hilbert state space (frequently in Fock space). Quantum mechanics and QFT make accent on the non-commutative nature of the dynamical variables but the interaction between pointwise particles and the relativistic invariance are borrowed from the classical theory. These are the sources of the singular functions involved in QFT. All attempts to build realistic extended stable model of an elementary particles in the framework of the relativistic linear QFT were failed. But it is clear that space-time in itself cannot be used in the consistent microscopic theory as an `arena' because the space-time localization of quantum particles is questionable \cite{W,Heg}.
It seems that just space-time properties should be established in some approximation to quantum dynamics {\it a posteriori}. Therefore new symmetries principle of the quantum state space should be used instead. New QFT will be based on this manifold with corresponding state coordinates. These coordinates are defined by the choice of some {\it quantum setup}. I will study the invariant relationships between quantum setups, say $S_A$ and $S_B$ as well as the relationships between the 
scales and clocks in the special or in the general relativity.
One should to have some formal setup description in order to express the new kind of invariance in the state space being applied to the state coordinates. 
The ordinary construction of the Hamiltonian describing some quantum setup $\hat{H}=\hat{H}_{particles}+\hat{H}_{fields}+\hat{H}_{int}$
is unsatisfactory in some reasons. First of all the separation between particle and its field is artificial. In fact the notion of particle in quantum theory is awkward. Second, the interaction term based on the classical energy of interaction, whereas it should appear as a consequence of quantum interaction. Third, the quantum motion requires the adequate mechanism of conservation or identification of quantum particles since the classical identification connected with the motion along space-time trajectory cannot be used.

In the present article I propose a non-linear relativistic 4-d field model originated by the internal dynamics in QPS. There is no initially distinction between `particle' and `field', and the space-time manifold is derivable. The main idea is to base the theory on the relative amplitudes solely. Quantum measurements
will be described in terms of parallel transport of the local dynamical variables and a specific gauge reduction of the full state vector to the Qubit coherent state. I will discuss here   field equations of quantum particle, arising in the dynamical space-time. Main specific points of my approach are listed below.

{\bf 1.} I introduce the operator of the Planck's action quanta $\hat{S}=\hbar {\hat \phi^+} {\hat \phi}$ with the spectrum $S_n=\hbar (n+1/2)$ in the separable Hilbert space $\cal{H}$. Since the action in itself does not create the gravity, it is legible to create the linear superposition $|\Psi>=\sum_{n=0}^\infty \Psi^n |n>$ of the action states where $\{|n>\}_0^{\infty}$ are $SU(\infty)$ multiplete of an abstract ``angular momentum''. The standard basis $|n,\phi>=(n!)^{-1/2} ({\hat \phi^+})^n|0>, n=0,1,2...$ corresponding to the discrete spectrum of the action operator $\hat{S}$ will be used. Because the notion of the quantum particles is awkward, I will discuss these superposition states and their dynamics, which leads to some non-linear field equations. These equations should presumably have the particle-like solutions.

{\bf 2.} Quantum state (the state of motion i.e. a {\it process} in the quantum setup \cite{Dirac1}) depends on the setup as well as the trajectory of a material point depends on the reference frame. Since setup and the quantum motions (quantum states in question) are, in fact, non-distinguishable, {\it there is the general inspiration to find the properties, invariant under the setup variation} (the super-relativity concept) \cite{Le1,Le2,Le3}. I will assume that invariant description is based {\it on the variation principle of the finite action closest to some initially chosen action}. The last is frequently given by the action functional of some classical model (the classical analogy method).

In order to formalize the non-local notion of the `quantum motion in setup A', I introduce the extremal $|\Psi_A>$ of the variation problem for the action functional $S[L_A]=\int_{t_0}^{t_1}dt \int_{x_0}^{x_1}{\cal{L_A}}(t,{\bf x})d^3x$. It means that temporary we (breaking the consistency) should preserve ordinary space-time coordinates in each setup as a reflection of the `coarse graining' \cite{Jones2}. Then the state vector $|\Psi_A>$ is some linear superposition of the functional basis $|\Psi_A>=\sum_{n=0}^\infty \Psi^n_A |n,\phi>$, where $\Psi^n_A=<n|\Psi_A>$. This vector is the generalized coherent state (GCS) of the $SU(\infty)$ action. Formally GCS and the local reference frame intrinsically connected with GCS (see below) replaces the vague notion of `setup'. Some different setup, say, $B$, will be described by the action
$S[L_B]=\int_{t'_0}^{t'_1}dt \int_{x'_0}^{x'_1}{\cal{L_B}}(t',{\bf x'})d^3x'$ where between coordinates $(t,{\bf x})$ and $(t',{\bf x'})$ there is no the generic relations. Sometimes, however, setup $B$ may be treated as a variation of the setup $A$. 
Then the model Lagrangian ${\cal{L_A}}(t,{\bf x})$ and
its extremal $<n|\Psi_A>$ are merely the ``initial conditions'' in the state space. The special tangent vector fields $\Phi^i_{\alpha}$ \cite{Le1} give the local functional frame originated at the action extremal.
The tangent vectors representing velocities of the action functional $S_A$ variation correspond to the local Hamiltonian or other LDV. This Hamiltonian has universal form being expressed in the local projective coordinates. Its specialization reveals only after the introduction of the space-time coordinates.   
In particular, the transition from setup $A$ to $B$ induces the space-time coordinates transformations involved the state space coordinates dependence. One should establish these transformations. The key idea is as follows. 

The behavior of the scales and clocks depends on the local gravitation potential (space-time curvature). In quantum theory the stationary orbits and self-frequencies, e.g. in atom, depend 
on the setup. Therefore the local dynamical variables at different GCS are state dependent because two different setups are in fact two different field (potential) configurations (different GCS of the action functional). 
Thus the quantum dynamics in the state space is similar to general relativity dynamics, where due to the equivalence principle, gravity is locally non-distinguishable from the accelerated reference frame. 
But in general relativity one has the distinction (by definition) between gravity (curvature) and its `matter' source. In quantum physics, however, all fields constituting setup and its quantum motion are `matter'. Variation of the setup fields transforms GCS due to interaction and the group structure of the setup transformations is given by the 

POSTULATE 1.

\noindent {\it Super-equivalence  principle: the unitary transformations of the action GCS may be identified with the action of physical unitary fields. The variation of the setup leads to the functional basis variation. These variations are generating by the global unitary transformations ${\hat U} \in G=SU(\infty)$ non-effectively acting on the GCS rays because the presence of the isotropy group $H=U(1)\times U(\infty)$ of some GCS $|\Psi>$. The result of the coset transformation $G/H=SU(\infty)/S[U(1)\times U(\infty)]=CP(\infty)$ is equivalent some physically distinguishable variation of GCS in $CP(\infty)$}. 

Therefore the reason for changing of GCS is the action of the state-dependent coset transformations $G/H=SU(\infty)/S[U(1)\times U(\infty)]=CP(\infty)$ on the rays of states in ${\cal H}$.
Thus these transformations being applied to the generalized coherent states are the quantum analog of classical forces acting on material points \cite{Le1}. 
In the framework of my model I assume that isotropy group of the vacuum $H=U(1)\times U(\infty)$ takes the place of the gauge group of `electroweak +' fields. Spontaneously broken $SU(\infty)$ symmetry of GCS leads to the motion in $CP(\infty)$ and therefore the local coordinates 
$(\pi^1=\frac{\Psi^1}{\Psi^j},...,\pi^k=\frac{\Psi^k}{\Psi^j},...)$ capable to specify any GCS. The choice of the map $\Psi^j \neq 0$ means, that the comparison of quantum amplitudes refers to the action $\hbar (j+1/2)$. Hence, all processes whose action differs considerably from 
$\hbar (j+1/2)$ will be taken into account in reduced measure due to the specific geometry of $CP(\infty)$ in comparison with `flat' Hilbert space. 
Thus $CP(\infty)$ manifold takes the place of the ground states (`local vacua'). 

{\bf 3.} Events realized by the quantum transitions in some two-level quantum subsystem treated as a notional `detector' giving `yes' or `no' answers under the measurement of the ``logical spin  1/2'' \cite{Le1} or Qubit coherent state. The state space of this Qubit is $C^2$ where acts $SU(2)$ sub-group of $SU(\infty)$ and its generalized coherent states is $CP(1)$. Thereby the space-time arises as a `section' of the state space $CP(\infty)$. The local Lorentz symmetry has priority over space-time coordinates. This invariance is now only the `two-level approximation' to the true $SU(\infty)$ symmetry. Therefore instead of the Lorentz group representation in some Hilbert space, I use opposite scheme of the `inverse representation' of $SU(\infty)$ or $SU(N)$ in $SU(2)$.

{\bf Local dynamical variables.} ---
The state space ${\cal H}$ of the field configurations with finite action quanta is a stationary construction. I introduce dynamics {\it by the velocities of variation of the GCS extremal} representing some `elementary quantum systems' (quantum particles). Their dynamics is specified by the Hamiltonian, giving velocities of variation of the action quanta number in different directions of the tangent Hilbert space $T_{(\pi^1,...,\pi^k,...)} CP(\infty)$ where takes place ordinary linear quantum scheme. The temp of the action variation gives energy of the particles.
 In fact only finite, say, $N$ action quanta are involved. Then one may restrict full QPS to finite dimensional $CP(N-1)$. 

The dynamical variables corresponding symmetries of the GCS and their breakdown should be expressed now in terms of the local coordinates $\pi^k$. The Fubini-Study metric
\begin{equation}
G_{ik^*} = [(1+\kappa \sum |\pi^s|^2) \delta_{ik}-\kappa \pi^{i^*} \pi^k](1+\kappa \sum |\pi^s|^2)^{-2}
\label{FS}  
\end{equation}
and the affine connection
\begin{eqnarray}
\Gamma^i_{mn} = \frac{1}{2}G^{ip^*}
(\frac{\partial G_{mp^*}}{\partial \pi^n} +
\frac{\partial G_{p^*n}}{\partial \pi^m})  
= - \kappa \frac{\delta^i_m \pi^{n^*} + \delta^i_n \pi^{m^*}}{1+\kappa \sum |\pi^s|^2}
\label{Gamma}  
\end{eqnarray}
in these coordinates will be used. Here $\kappa=R^{-2}$ is the curvature of the sphere serving as a model of $CP(N-1)$ through the stereographic projection. I will assume temporary that $R=1$ for simplicity.
Hence the internal dynamical variables and their norms should be state-dependent, i.e. local in the state space \cite{Le1,Le2,Le3,Le4}. These local dynamical variables realize the non-linear representation of the unitary global $SU(N)$ group in the  Hilbert state space $C^N$. Namely, $N^2-1$ generators of $G = SU(N)$ may be divided in accordance with Cartan decomposition: $[B,B] \in H, [B,H] \in B, [B,B] \in H$. Namely,  $(N-1)^2$ generators are the generators 
$\Phi_h^i \frac{\partial}{\partial \pi^i}+c.c. \in H,\quad 1 \le h \le (N-1)^2 $ of the isotropy group $H = U(1)\times U(N-1)$ of the ray (Cartan sub-algebra) and $2(N-1)$ generators 
$\Phi_b^i \frac{\partial}{\partial \pi^i} + c.c. \in B,\quad 1~ \le b \le 2(N-1)$ are the coset 
$G/H = SU(N)/S[U(1) \times U(N-1)]$ generators realizing the breakdown of the $G = SU(N)$ symmetry disturbing GCS. 
Furthermore, $(N-1)^2$ generators of the Cartan sub-algebra may be divided into the two sets of operators: $1 \le c \le N-1$ ($N-1$ is the rank of $Alg SU(N)$) Abelian operators, and $1 \le q \le (N-1)(N-2)$ non-Abelian operators corresponding to the non-commutative part of the Cartan sub-algebra of the isotropy (gauge) group.
Here $\Phi^i_{\sigma}, \quad 1 \le \sigma \le N^2-1 $ are the coefficient functions of the generators of the non-linear $SU(N)$ realization. They give the infinitesimal shift of $i$-component of the coherent state driven by the $\sigma$-component of the unitary multipole field rotating the generators of $Alg SU(N)$ and defined as follows:

\begin{equation}
\Phi_{\sigma}^i = \lim_{\epsilon \to 0} \epsilon^{-1}
\biggl\{\frac{[\exp(i\epsilon \lambda_{\sigma})]_m^i \Psi^m}{[\exp(i \epsilon
\lambda_{\sigma})]_m^j
\Psi^m }-\frac{\Psi^i}{\Psi^j} \biggr\}=
\lim_{\epsilon \to 0} \epsilon^{-1} \{ \pi^i(\epsilon \lambda_{\sigma}) -\pi^i \},
\end{equation}
\cite{Le1}. Then the sum of $N^2-1$ the energies of `elementary systems' (particle plus fields) is equal to the excitation energy of the GCS, and the local Hamiltonian $\vec{H}$ is linear against the partial derivatives $\frac{\partial }{\partial \pi^i} = \frac{1}{2} (\frac{\partial }{\partial \Re{\pi^i}} - i \frac{\partial }{\partial \Im{\pi^i}})$
and 
$\frac{\partial }{\partial \pi^{*i}} = \frac{1}{2} (\frac{\partial }{\partial \Re{\pi^i}} + i \frac{\partial }{\partial \Im{\pi^i}})$, i.e. it is the tangent vector to $CP(N-1)$
\begin{equation}
\vec{H}=\vec{V_b}+\vec{T_c}+\vec{T_q}= \hbar \Omega^b \Phi_b^i \frac{\partial }{\partial \pi^i} +
\hbar \Omega^c \Phi_c^i \frac{\partial }{\partial \pi^i} + 
\hbar \Omega^q \Phi_q^i \frac{\partial }{\partial \pi^i} + c.c.. 
\label{field}
\end{equation}

Since $T_c=\hbar \Omega^c \Phi_c^i = \hbar  \sum_{c=1}^{N-1} \Omega^c \alpha^i_{k c} \pi^k$, where $\alpha^i_{k c}$ are constants, this term may be identify with the energy of the independent (commutative) set of harmonic oscillators (`photons' of my model). 
Operators $\Phi_q^i \frac{\partial }{\partial \pi^i}$ are the non-commutative components of the isotropy sub-group and corresponds to some non-Abelian gauge fields. Their sum is the Hamiltonian 
${\vec T}=T^i \frac{\partial }{\partial \pi^i} + c.c.$ of the self-conservation of GCS. The coset generators correspond to the perturbation energy 
${\vec V}=V^i \frac{\partial }{\partial \pi^i} + c.c.$ trying to spread GCS. The competition of these energies defines the fate of the GCS. The eigen-problem for this Hamiltonian
$\vec{H} F =\Lambda F$ may be classified accordantly to decomposition mentioned above. In the simplest case of the constant coefficient $\Omega^{\alpha}$ such equations may be solved analytically. For example the real part of the quasi-linear equation in the partial derivatives 
$\vec{T_c} F =\Lambda_c F$ defines the quasi-homogeneous Euler field. The imaginary part gives the curl field as a hint on the intrinsic structure of the extended quantum particles.  
Their characteristic equations give the parametric representations of the solutions in $CP(N-1)$. The parameter $\tau$ in these equations I will identify with ``universal time of evolution'' of Newton-Stueckelberg-Horwitz-Piron (NSHP) \cite{H1}. 

{\bf Field equations in the dynamical space-time.}---
Up to now we discuss internal dynamics of the LDV. We should embed this dynamics in dynamical space-time. What is the space-time from the quantum point of view? 
How its structure may be established using complex amplitudes solely? These questions should recall reader to look in the depth of our notion about space-time. 
The space-time arises as a result of the neglect of the gigantic number of the degrees of the freedom of observable objects. Thus the notion of the material point lays in the base of the space-time geometry. Collisions of material points are events. 
Only the invariant properties of the events positions relative some reference frame, are essential for the space-time structure.
Shortly speaking it is based on the abstraction of the coincidence of two events. The conclusion about coincidence is based on the detector click as a reply on the `question': `yes' or `no'. The physically invariant (macroscopic) structure of their coincidence is the pseudo-Euclidean space-time geometry of Einstein \cite{Einstein1}. The general relativity, however, demolished this stiff global structure \cite{Einstein2}.
In quantum theory one has much more difficult situation since we lose the notion of the material point. Furthermore, we have the localization problem of the wave function. Its different aspects have been studied particulary in works \cite{W,Heg}. In order to avoid these problems I will use the LDV comparison due to the parallel transport in $CP(N-1)$ and the full GCS squeezing procedure up to the Qubit coherent state. This is, unfortunately, complicated procedure.     

Let me assume that GCS is $|\Psi>=(\Psi^0,\Psi^1, ...,\Psi^N)^T$. The first stage of a measurement is the reduction of the full state vector $|\Psi>$ to the two-level state of a ``detector''.
This is the model of the neglect of the all degrees of freedom besides two degrees belonging to some notional ``detector''. The ``detector'' has, say, two states $|1>=(1,0,...,0)^T$ and $|0>=(0,1,0,...,0)^T$. Then the Qubit coherent state of the ``detector'' is $|Q>=(\cos \Theta,  e^{i f}\sin \Theta,0, ..., 0)^T$. If one assume that initial state $(1,0,...,0)^T$ should be intact under the setup tuning, only the isotropy group $U(1)\times U(N-1)$ is acceptable for the tuning. The last will be realized by the squeezing ansatz. One can render the GCS $|\Psi>$ into the ``two-level state'' as follows. The first ``squeezing'' unitary matrix is
\begin{equation}
\hat{G}_1^+= \left(
\matrix{
1&0&0&.&.&.&0 \cr
0&1&0&.&.&.&0 \cr
.&.&.&.&.&.&. \cr
.&.&.&.&.&.&. \cr
0&.&.&.&1&0&0 \cr
.&.&.&.&0&cos \phi_1&e^{i\alpha_1} sin \phi_1 \cr
0&0&.&.&0&-e^{-i\alpha_1}sin \phi_1&cos \phi_1 
}
\right ). 
\end{equation}
This matrix acts on the state vector $|\Psi>$ with the result
\begin{equation}
\hat{G}_1^+|\Psi>= \left(
\matrix{
\Psi^0 \cr
\Psi^1 \cr
. \cr
. \cr
. \cr
\Psi^{N-1} \cos \phi_1
+ \Psi^{N} e^{i\alpha_1} \sin \phi_1 \cr
-\Psi^{N-1} e^{-i\alpha_1}\sin \phi_1
+ \Psi^{N} \cos \phi_1 
}
\right ).
\label{G1} 
\end{equation}
Now one has solve two ``equations of annihilation'' \cite{Le1}
$\Re (-\Psi^{N-1} e^{-i\alpha_1}\sin \phi_1
+ \Psi^N \cos \phi_1 )=0$
and 
$\Im ( -\Psi^{N-1} e^{-i\alpha_1}\sin \phi_1
+ \Psi^N \cos \phi_1)=0$ in order to
eliminate the last string and to find $\alpha'_1$ and $\phi'_1$.
That is one will have a squeezed state vector
\begin{equation}
\hat{G}_1^+|\Psi>= \left(
\matrix{
\Psi^0 \cr
\Psi^1 \cr
. \cr
. \cr
. \cr
\Psi^{N-1} \cos \phi'_1
+ \Psi^N e^{i\alpha'_1}\sin \phi'_1\cr
0
}
\right ).
\label{G2} 
\end{equation}
The next step is the action of the matrix with the shifted transformation block
\begin{equation}
\hat{G}_2^+= \left(
\matrix{
1&0&0&.&.&.&0 \cr
0&1&0&.&.&.&0 \cr
.&.&.&.&.&.&. \cr
0&.&.&.&1&0&0 \cr
.&.&.&.&0&cos \phi_2&e^{i\alpha_2} sin \phi_2 \cr
0&0&.&.&0&-e^{-i\alpha_2}sin \phi_2&cos \phi_2 \cr
0&.&.&.&0&0&1 
}
\right )
\label{G3} 
\end{equation}
on the vector and the evaluation of
$\alpha'_2$ and $\phi'_2$ and so on till the initial 
state vector $|\Psi>$ will be reduced to the Qubit spinor 
 $|Q(\Psi^0,...,\Psi^N)>= (\Psi^0, e^{i\Lambda(\Psi)} \Psi^1, 0,...,0)^T=\hat{G}_{N-2}^+...\hat{G}_1^+ |\Psi>$.
It is the Qubit coherent state of the two-level ``detector'' 
arising due to the squeezing procedure during the unitary setup tuning. It is clear that the choice of the ``detector'' and the squeezing procedure is subjective (observer involved), therefore this is not single defined. 

Formally the two infinitesimally close Qubit coherent states $|Q1>=(\cos(\Theta)e^{i \psi},\sin(\Theta)e^{i (\Lambda+\psi})^T$
and
$|Q2>=(\cos(\Theta+\epsilon_1)e^{i (\psi+\epsilon_3)},\sin(\Theta+\epsilon_1)e^{i (\Lambda+\epsilon_2+\psi+\epsilon_3)})^T$ may be connected by the unitary matrix
\begin{equation}
\hat{U}_L=\left(
\matrix{
L_{11}& L_{12}\cr
L_{21}& L_{22}
}
\right ),
\end{equation}
with the elements $L_{11}=1-i(\epsilon_2 \sin^2 \Theta+\epsilon_3 \cos 2\Theta)$,
$ L_{12}=-\epsilon_1\cos \Lambda +(1/2) \epsilon_2\sin 2\Theta \sin \Lambda+\epsilon_3 \sin 2 \Theta \sin \Lambda +i(\epsilon_1\sin \Lambda +(1/2) \epsilon_2\sin 2\Theta \cos \Lambda+\epsilon_3 \sin 2 \Theta \cos \Lambda) $, $L_{21}=-L_{12}^*$, $L_{22}= L_{11}^*$.
But in order to establish the {\it fact of the coincidence in our ``detector'' the expected particle with caught particle}, each infinitesimally close transformation $|Q1>$ into $|Q2>$ should be accompanied with small parallel transport of the Hamiltonian tangent vector field. Then the small shifts $\epsilon_1, \quad \epsilon_2$ give the direction of the parallel transport of the Hamiltonian in $CP(1)$ whereas $\epsilon_3$ is the angle of the Hamiltonian `rotation' during the parallel transport.

Objectively LDV are parallel transported in  $V^k$ direction of $CP(N-1)$ during the NSHP evolution but the comparison of this natural LDV with an expectation value of the LDV in two close Qubit states is the subjective measurement process of the `field of the particle' revealed in the dynamical space-time. Now we should to get the field equations.  

The problem of the identification of pointwise quantum particles has been discussed in terms of the dense measurement \cite{AV}. 
Extended quantum particles (excitations of GCS represented by tangent vectors to $CP(N-1)$) should be identified during NSHP evolution on the base of the affine parallel transport of some LDV in $CP(N-1)$ \cite{Le4}. Note that parallel transport comprises the continuous projection onto the tangent space at all intermediate GCS's. But this parallel transport has merely blameless geometric content without the space-time field realization. The local gauge nature of this parallel transport (affine connection in $CP(N-1)$) will be revealed due to the super-relativity principle. Namely, {\it in accordance with the super-equivalence one should to find such infinitesimal shift of the unitary field} $\delta \Omega^{\alpha}$ {\it in the dynamical space-time that gives the infinitesimal shift of the energy, equivalent to the infinitesimal shift of tangent Hamiltonian field generated by the parallel transport in} $CP(N-1)$ {\it during NSHP time} $\delta \tau)$. Thus one has 
$\hbar (\Omega^{\alpha} + \delta \Omega^{\alpha} ) \Phi^k_{\alpha}=
\hbar \Omega^{\alpha}( \Phi^k_{\alpha} - \Gamma^k_{mn} \Phi^m_{\alpha} V^n \delta \tau)$
and, hence,
$\frac{ \delta \Omega^{\alpha}}{\delta \tau} = - \Omega^{\alpha}\Gamma^m_{mn} V^n $.

I will introduce dynamical space-time coordinates as numbers,
transforming in accordance with the local ``Lorentz transformations'' 
$x^{\mu} + \delta x^{\mu} = (\delta^{\mu}_{\nu} + \Lambda^{\mu}_{\nu} \delta \tau )x^{\nu}$. The parameters of $\Lambda^{\mu}_{\nu} (\Psi)$ arise as a result of the global gauge transformations 
in $CP(N-1)$ of the squeezing full state vector up to the coherent state of the Qubit (the tuning process of the setup).
Physically it means that the global infinitesimal gauge transformations and the last step of the coset transformation of the quantum setup may be compensated by the infinitesimal Lorentz transformations 
$\hat{L}=1+\frac{1}{2}\tau \vec{\sigma}(\vec{a}-i\vec{\omega})$
(particle acquires an additional energy-momentum).
Then $\frac{\delta x^{\mu}}{\delta \tau} = \Lambda^{\mu}_{\nu} (\Psi) x^{\nu} $ and therefore the energy balance equation reads now
\begin{equation}
\Lambda^{\mu}_{\nu} (\Psi) x^{\nu} \frac{\partial \Omega^{\alpha}}{\partial x^{\mu} } = - \Omega^{\alpha}\Gamma^m_{mn} V^n.
\end{equation}
If one wish find the field corresponding to the given
trajectory, say, a geodesic in $CP(N-1)$, then, taking into account that any geodesic as whole lays in  some $CP(1)$, one may put $ \pi= e^{i\gamma} \tan(\omega \tau)$. Then $V=\frac{d \pi}{d \tau}=\omega \sec^2(\omega \tau) e^{i\gamma}$, and one has the linear wave equations for the gauge unitary field $\Omega^{\alpha}$ in the dynamical space-time with complicated coefficient functions of the local coordinates $(\pi^1,...,\pi^{N-1})$. In the case of the spherical symmetry $(x^0=ct,x^1=r\sin \Theta \cos \Phi, x^2=r\sin \Theta \sin \Phi, x^3=r\cos \Phi)$, under the assumption $\tau = \beta t$ this system has following solution
\begin{eqnarray}
\Omega^{\alpha}=(F_1^{\alpha}(r^2-c^2t^2)+i F_2^{\alpha}(r^2-c^2t^2)) \exp{(8\beta c \int dp \frac{\sin(\beta p)}{D(p) \sqrt{c^2(p^2-t^2)+r^2}})} ,
\end{eqnarray}
where $D(p)=a_1(\sin(\beta p +\Theta +\Phi)+ \sin(-\beta p +\Theta +\Phi)+ \sin(\beta p +\Theta -\Phi)+ \sin(\beta p +\Theta -\Phi))+a_2(\cos(-\beta p +\Theta -\Phi)+ \cos(\beta p +\Theta -\Phi)- \cos(-\beta p +\Theta +\Phi)-\cos(\beta p +\Theta +\Phi))+2a_3(\cos(-\beta p +\Theta)+ \cos(\beta p +\Theta))$.
It is interesting that the general factor demonstrates the diffusion of the light cone (mass shell) due to the boosts. 
Thereby, so-called ``off-shell'' ideology by Horwitz \cite{H2}, got now the quantum support. Namely, boost parameters $\vec{a}$ and angle velocities $\vec{\omega}$ one may obtain due to the comparison of two matrices: the infinitesimal Lorentz transformations $\hat{L}$ and the unitary matrix $\hat{U}_L$. 
All these parameters rooted in the $CP(N-1)$ geometry and depend on the whole prehistory of the initial GCS $|\Psi>$ due to squeezing procedure describing the tuning of the setup. 

The self-consistent problem 
\begin{equation}
\Lambda^{\mu}_{\nu} (\Psi) x^{\nu} \frac{\partial \Omega^{\alpha}}{\partial x^{\mu} } = - (\Gamma^m_{mn} \Phi_{\beta}^n+\frac{\partial \Phi_{\beta}^n}{\partial \pi^n}) \Omega^{\alpha}\Omega^{\beta}, \quad \frac{d\pi^k}{d\tau}=\Phi_{\beta}^k \Omega^{\beta} 
\end{equation}
arising under the condition of the parallel transport of the Hamiltonian field $H^k=~\Phi^k_{\alpha} \Omega^{\alpha}$
is more interesting and much more difficult. 
The right part consists of the square non-linearity in fields $\Omega^{\alpha}$ capable to give presumably the envelope solitons. It would be interesting to analyze these solutions. 
The self-consistent time evolution of the two-level system was studied in \cite{Le5}. Complicated self-consistent system of eight ordinary non-linear differential equations of the parallel transport unfortunately cannot be solved analytically. Hence we dispose only numerical solutions for $\Omega^\alpha$ and $\pi$
with interesting periodical and quantal solutions. A more detailed analysis will be necessary for the self-consistent problem mentioned above.   

{\bf Conclusions}---

1. Dynamical $3+1$ space-time arises naturally under the unitary squeezing the superposition state up to the coherent state of the Qubit, has locally pseudo-Euclidean structure. The curvature of $CP(N-1)$ may lead globally to the pseudo-Riemannian structure of the dynamical space-time. Seems to me it may pave the way to the consistent quantum gravity.   

2. The result of the squeezing looks like indeterminate  (probabilistic) since the squeezing process is case-sensitive (observer involved) and it is frequently out of full our control. 
Therefore, the microscopic (quantum) origin of the space-time and the indeterminate character of the quantum motion in space-time are closely connected.  

3. The decoherence arises naturally, since only action GCS (action waves) may keep up linear superposition. But energy packets (tangent vectors to $CP(N-1)$ should gravitate, hence their superposition may be linear only in some approximation.

\vskip 0.2cm
I am grateful to L.P.Horwitz for a lot of interesting discussions and critical notes. Further, I thank my wife for her patience during many years.


\vskip 0.2cm 
\begin{thebibliography}{99}
\bibitem{Jones1}
K. R. W. Jones, Ann. Phys. (N.Y.), {\bf 233}, 295 (1994).
\bibitem{Jones2}
K. R. W. Jones, Aust. J. Phys., {\bf 48}, 1055 (1995).
\bibitem{CMP}
R.Cirelli, A.Mani\'a, L.Pizzocchero, Int.J. of Mod. Phys. 
A {\bf 6}, (12) 2133 (1991). 
\bibitem{Hughston1}

L.P.Hughston, ``Geometry of Stochastic State Vector Reduction", Proc. R. Soc. Lond. {\bf A452}, 953 (1996).
\bibitem{Hughston2}
L.P.Hughston, ``Geometric Aspects of Quantum Mechanics", Chapter 6 in: {\it Twistor Theory} (ed. S.A.Huggett), pp.55-79. Marcel Dekker. 
\bibitem{Ashtekar}
A.Ashtekar and T.A.Schilling, arXive: gr-qc/9706069.
\bibitem{AdHr}
S.L.Adler, L.P.Horwitz, arXive: quant-ph/9909026. 
\bibitem{W}
T.D.Newton and E.P.Wigner, Rev. Mod. Phys., {\bf 21}, 400 (1949).
\bibitem{Heg}
G.C.Hegerfeldt, Phys.Rev.D{\bf 10}, 3320 (1974).

\bibitem{Dirac1}
P.A.M. Dirac, {\it The principles of quantum mechanics}, Oxford, Clarendon Press, 1958.
\bibitem{Le1}
P.Leifer, Found. Phys.{\bf 27}, (2) 261 (1997).
\bibitem{Le2}
P.Leifer, Found.Phys.Lett., {\bf 11}, (3) 233 (1998).
\bibitem{Le3}

P.Leifer, arXive: gr-qc/0201039.

\bibitem{Le4}
P.Leifer, JETP Letters, {\bf 80}, (5) 367 (2004).
\bibitem{H1}
L.P.Horwitz, hep-ph/9606330

\bibitem{Einstein1}
A.Einstein, Ann. Phys. {\bf 17}, 891 (1905).
\bibitem{Einstein2}
A.Einstein, Ann. Phys. {\bf 49}, 769 (1916).
\bibitem{AV}
Y.Aharonov, M.Vardi, Phys.Rev.D{\bf 21}, (8) 2235 (1980).

\bibitem{H2}
J.Frastai and L.P.Horwitz, Found. Phys. {\bf 25}, (10) 1495 (1995).
\bibitem{Le5}
P.Leifer, Found.Phys.Lett., (in press).


\end{thebibliography}
\end{document}